\documentclass{iaus}
\usepackage{graphicx}

\title[Kinetic Luminosity of AGN] 
{Cosmological evolution of the AGN Kinetic
  Luminosity Function}

\author[Merloni and Heinz]   
{A. Merloni$^1$, S. Heinz$^2$}
\affiliation{$^1$Max Planck Institut f\"ur Astrophysik, Garching, Germany\break email:
  am@mpa-garching.mpg.de\\[\affilskip]
$^2$ Department of Astronomy,
University of Wisconsin, Madison, WI, USA}
\pubyear{2006}
\volume{238} 
\pagerange{001--999}
\date{??? and in revised form ???}
\jname{Black Holes: from Stars to Galaxies -- across the Range of
  Masses}
\editors{V. Karas \& G. Matt, eds.}
\begin{document}
\maketitle

\begin{abstract}
We present a first attempt to derive the cosmological evolution of the kinetic
luminosity function of AGN based on the joint evolution of the flat
spectrum radio and hard X-ray
selected AGN luminosity functions. An empirical correlation between
jet power and radio core luminosity is found, which is consistent with
the theoretical assumption that, below a certain Eddington ratio, SMBH
accrete in a radiatively inefficient way, while most of the energy
output is in the form of kinetic energy. We show  how the redshift
evolution of the kinetic power density from such a low-$\dot m$ mode
of accretion makes it a good candidate to explain the so-called
``radio mode'' of AGN feedback as postulated in many galaxy formation schemes.  
\end{abstract}
\firstsection 
\section{Introduction: the radio mode of AGN}
Supermassive black holes in the nuclei of many
nearby bulges are fundamental constituents of their
parent galaxies. The observed correlations between black
hole mass and large scale bulge properties (Marconi \& Hunt
2003, and references therein) 
suggest there must have been an epoch in which the central 
black hole had a direct impact on the galaxy's
structure. Recent sustained theoretical 
efforts have indeed identified the period of most
rapid  black hole growth as the most likely epoch when such an
impact was more dramatically felt. 
As it is now well established that most of this growth is
to be identified with episodes of accretion, when galactic nuclei
become ``active'' (Yu and Tremaine 2002; Marconi et al. 2004), 
AGN feedback is regarded as a
necessary ingredient in all models of galaxy formation and evolution.

In particular, numerical simulations of merging galaxies with
SMBH growth and radiative feedback (Di Matteo et
al. 2005) demonstrated the
effectiveness of quasars in quenching subsequent episodes of star
formation.
However, semi-analytic schemes of galaxy formation (Croton et al. 2006;
Bower et al. 2006) as well as indirect
evidence of AGN-induced heating in clusters of galaxies, both seem to
indicate that a different mode of accretion onto SMBH, not directly
linked to bright quasar phases (and therefore termed ``radio mode'')
must be effective in the most massive system at
late times in order not to over-predict the observed abundances 
of high mass galaxies
as well as to explain the absence of vigorous cooling flows in the center of 
clusters.

\section{A simple scaling for the jet kinetic energy}
A mode of AGN feedback dominated by kinetic energy 
is in fact expected on theoretical grounds (Rees et al. 1982) 
in black holes of low luminosity (in units of
the Eddington one $L_{\rm Edd}=1.3 \times 10^{38} M/M_{\odot}$ erg
s$^{-1}$). Observational evidence for such ``jet dominated'' 
modes of accretion come
from either black holes of stellar mass in the so-called low/hard or
quiescent states
(Fender, Gallo \& Jonker 2003; Gallo et al. 2006) and low-luminosity
AGN (Merloni, Heinz
and Di Matteo 2003 [MHD03]; Falcke, K\"ording \& Markoff 2004). In both cases,
the observed correlations between the (unresolved) radio luminosity of
the jet core and its X-ray luminosity can be interpreted as a
by-product of the coupling between a radiatively inefficient accretion
flow and a self-absorbed (quasi-conical) jet, provided that the
kinetic luminosity of the jet obeys $L_{\rm kin} \propto \dot M_{\rm
  out} c^2$, where $\dot M_{\rm out}$ is the accretion rate at the
outer boundary of the accretion flow, roughly coincident with the
Bondi radius in the case of AGN.
The constant of proportionality in the above relationship cannot be
directly inferred from the observed radio-X-ray correlations, but has
to be determined by direct measurements of jet kinetic energy.
On the basis of a very sparse collection of measurements of this sort,
Heinz and Grimm (2005) originally proposed the
following scaling: 
\begin{equation}
\label{eq:scal}
L_{\rm kin}=6 \times 10^{37} (L_{\rm R}/10^{30})^{0.7} {\rm erg/s,}
\end{equation}
where $L_{\rm R}$ is the radio luminosity of the core at 5 GHz
measured in ergs s$^{-1}$. In fact, it can be shown
that such a normalization corresponds to an extreme case in
 which essentially all accretion power $\eta \dot M_{\rm out} c^2$, for
 $\eta\simeq 0.1$, is released as kinetic energy, on the basis of the
 observed ``fundamental plane'' relation of MHD03.

\begin{figure}
 \centerline{
\scalebox{0.33}{\includegraphics[angle=270]{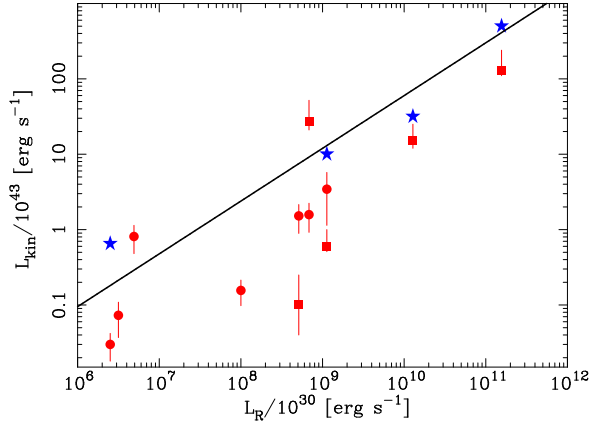}}}
  \caption{Estimated jet kinetic luminosity as a function of the 5GHz core
    radio luminosity. Stars represent direct kinetic energy estimates
    based on radio lobe modeling for NGC 4636, M87, Per A and Cyg A, in order
    of increasing power. Filled circles are the kinetic energy
    estimates from Allen et al. (2006), filled squares those from
    Rafferty et al. (2006). The solid line is the theoretical scaling
  $L_{\rm kin}=6 \times 10^{37} (L_{\rm R}/10^{30})^{0.7}$.}\label{fig:power}
\end{figure}

One way to test the above relation is to look for alternative,
indirect ways to estimate the kinetic power of the AGN jet. This has
been done recently by Allen et al. (2006) and Rafferty et
al. (2006) by measuring $PdV$ work done by the
jets on the intracluster gas. In Figure~\ref{fig:power} we show these
observational data points, together with a few estimates of kinetic power
from direct modeling of the radio lobe emission (stars) plotted as a
function of the radio core luminosity, either measured directly or,
preferentially, from the black hole mass and 2-10 keV nuclear
luminosity and the MHD03 relation (see Heinz \& Grimm 2005 for
references). 
In the same figure we also plot relation (\ref{eq:scal}). Indeed,
the observational points do not scatter too far away from the
predicted relation, even more so if one considers that the kinetic
power measurements from X-ray cavities (circles and squares), although
averages,  
are most likely lower limit to
the intrinsic jet power, due to the additional presence of weak shocks
and low-contrast cavities (Nulsen et al. 2006).

\section{The kinetic luminosity evolution}
With the aid of eq. (\ref{eq:scal}), we are now in the position 
to attempt a measurement of the kinetic luminosity
function of AGN. 
First, we need to derive the intrinsic 5GHz jet core luminosity
function of AGN. This can be done by analytically de-beaming
the observed luminosity function of flat spectrum radio quasars
and blazars (dominated by relativistically beamed sources), 
via the formalism described in
detail in Urry and Padovani (1991). The analysis is greatly simplified
if a narrow distribution of Lorentz factor is assumed for the jets, and, for
the sake of simplicity, we assume this holds true here. 

\begin{figure}

\centerline{
\scalebox{0.37}{\includegraphics{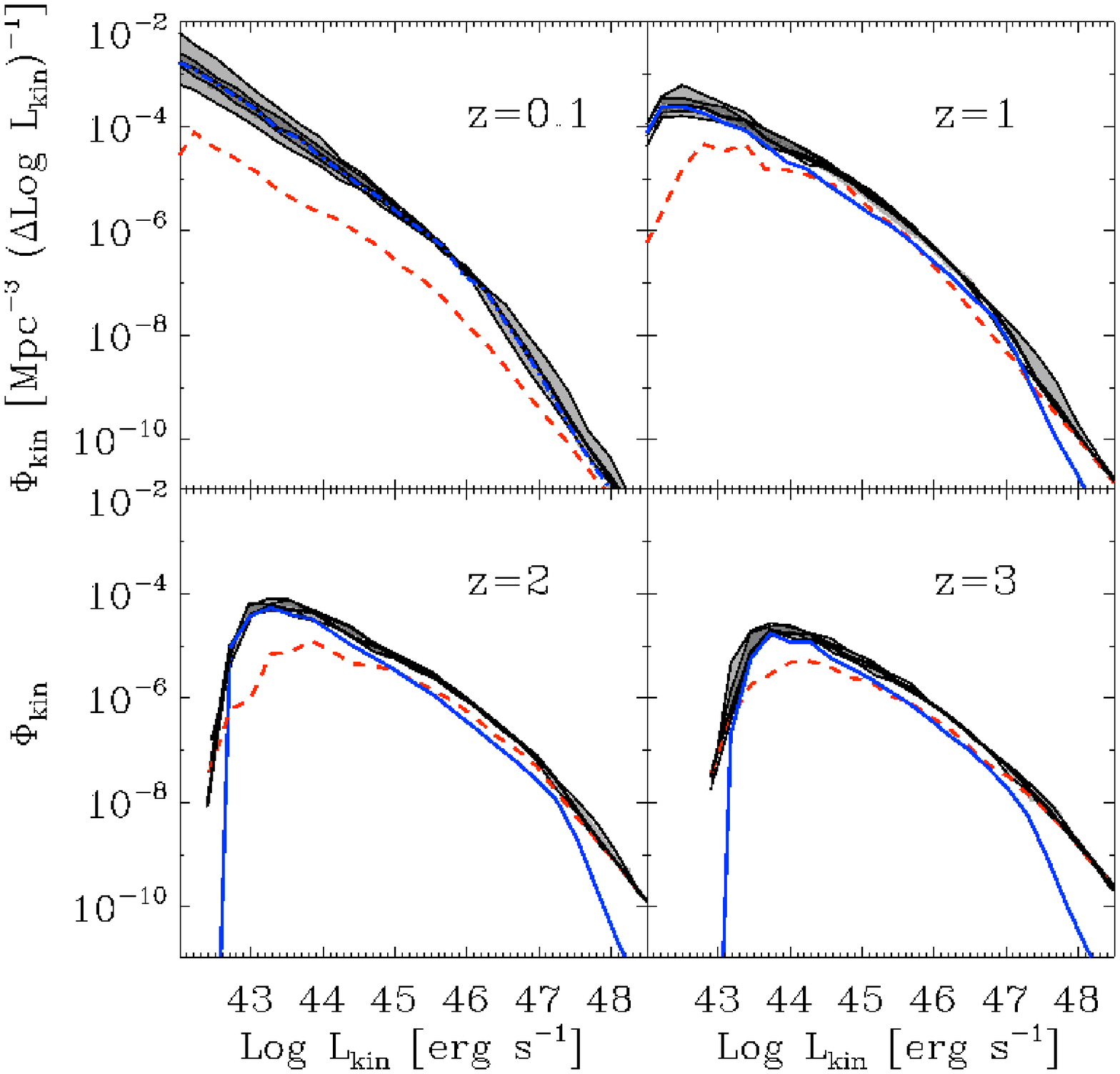}}\hspace{-0.65cm}\scalebox{0.36}{\includegraphics{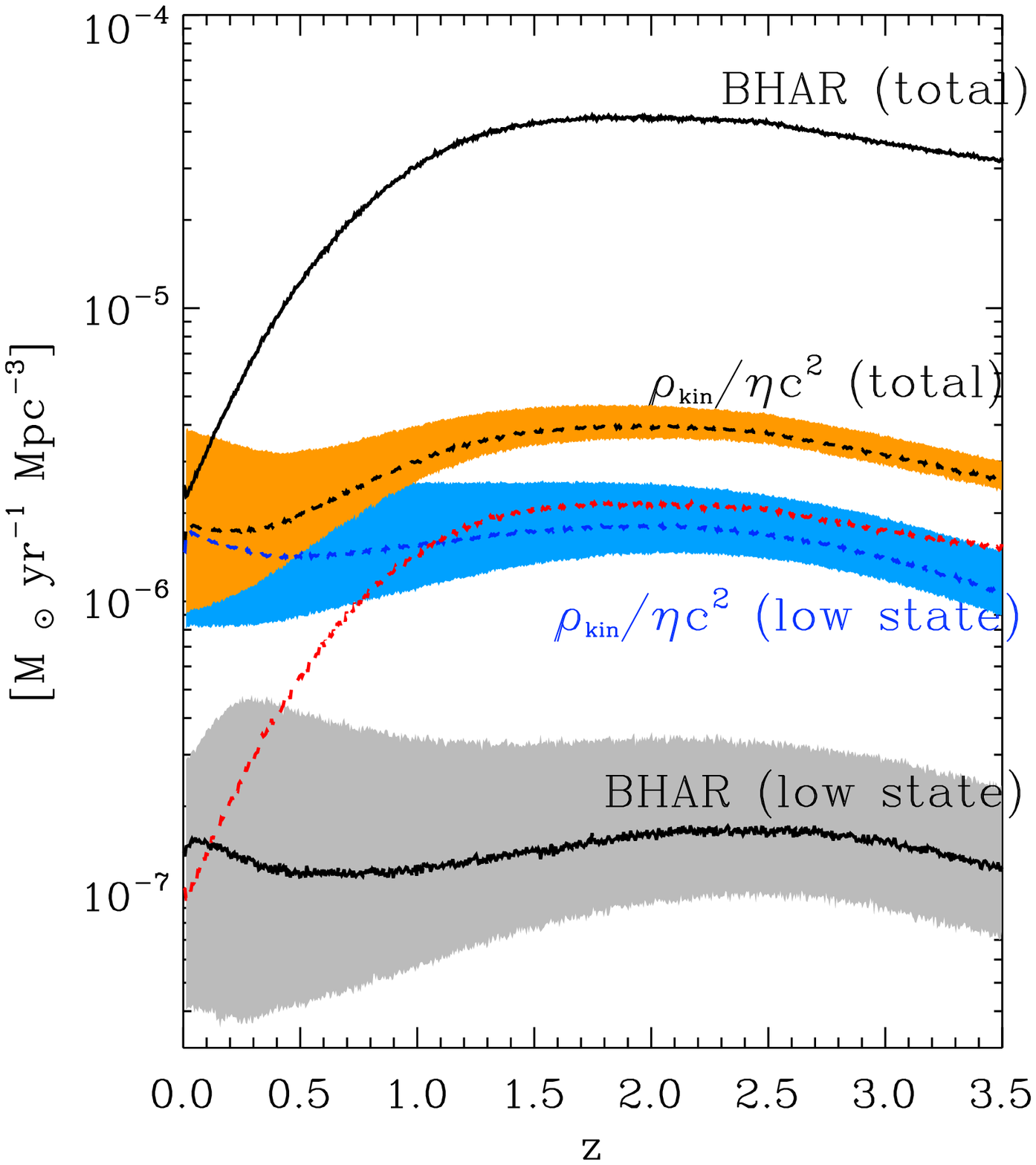}}}
  \caption{\textit{Left panel:} Total
    kinetic luminosity function (solid lines) at various redshifts as
     the sum of low-
    and high-$\dot m$ modes (solid blue and dashed red,
    respectively). The shaded areas highlight the uncertainties due to
variation of the average jets' Lorenz factors ($2<\Gamma<10$, 
black solid line corresponds to
$\Gamma=5$). The low luminosity downturns are due
to incompleteness in the high-redshift radio and X-ray luminosity
functions adopted here. \textit{Right panel:} Redshift evolution of the
    kinetic energy 
    density mass equivalent (calculated assuming a energy conversion
    efficiency of $\eta=0.1$; black dashed line,
    orange shaded area),
decomposed into the sum of the low- (blue dashed line, cyan shaded area)
and high-$\dot m$ (red dashed line) modes, and compared with the mass
accretion rate onto the black holes (BHAR; black solid lines).}\label{fig:bhar}
\end{figure}

One can then use the technique developed in Merloni (2004) to unveil
the accretion history of SMBH based on the combined evolution of the
hard X-ray and radio cores luminosity function coupled with the
fundamental plane relationship.
Here, differently from Merloni (2004),
we adopt an intrinsic radio core luminosity
function evolution, based on the observed FSRQ/Blazar luminosity
function of De Zotti et al. (2005), to which we added local constraints
on the faint-end slope from Filho, Barthel \& Ho (2006). 
We adopt the scaling
(\ref{eq:scal}) for all sources accreting below a critical Eddington
fraction, that  we have fixed at about 3\% (see Merloni 2004). Such objects, by
construction, will lie on the fundamental plane of black hole
activity and represent the majority of radio sources below the
FRI/FRII divide. We assume here that only a small fraction of the
accreting matter (powering the jet) makes its way onto the
central black holes, i.e. we assume that black holes are always
radiatively efficient {\it with respect to the accreted gas}, rather
than to the liberated energy (see the discussion
in Gallo et al. 2006).
For object whose accretion
rate is above 3\%, we assume that the vast majority (90\%) are
radio quiet and radiatively efficient (with $\epsilon_{\rm rad}=\eta=0.1$),
while for the remaining 10\% the liberated energy is equally divided
into radiative and kinetic. Those powerful radio loud quasar dominate
the high end of the radio luminosity function (FR II)\footnote{It is worth noticing
here that the above recipe to include powerful radio sources is highly
arbitrary, pending a more complete knowledge of the physical nature of
powerful radio loud quasars. Therefore, all our conclusions regarding the
evolution of this class of sources, and of their kinetic energy output, should
be considered as indicative only.}. 

Given such a recipe for the
accretion modes (or, equivalently, for the radiative and kinetic
efficiencies as functions of $\dot m$), 
the local SMBH mass function can be evolved backwards in time 
according to the corresponding
accretion rate distribution, and the procedure
is reiterated until acceptable fits to the observed hard X-ray
and radio core luminosity functions are found.

Figure~\ref{fig:bhar} shows, on the left panel, the evolution of the
computed kinetic luminosity function of AGN at four different
redshifts. By integrating these luminosity functions one can obtain the
redshift evolution of the kinetic energy density, $\rho_{\rm kin}$, provided by
SMBH. This is shown in the right panel of fig.~\ref{fig:bhar}. Here the
uncertainties in $\rho_{\rm kin}$ include not only the jets' Lorentz factor 
distribution, but also
uncertainties on the faint end slope of the radio core
luminosity function. The redshift evolution of $\rho_{\rm kin}$
differs substantially from that of the accretion rate (BHAR) density, 
strongly resembling instead the required ``radio mode'' evolution of
 galaxy formation models
(Croton et al. 2006; Bower et al. 2006). 
A more detailed study of these results is underway and will be
presented elsewhere (Heinz et al., Merloni et al., in preparation).

\section{Conclusions}
A separate mode of SMBH growth in which most of the energy is released
in kinetic form is postulated both on theoretical grounds and from
very general arguments of galaxy formation theory. Here we have
discussed how 
direct observational evidence of such a mode emerging from recent
studies of X-ray binary black holes can be generalized to the case of
SMBH, when the accretion rate (in Eddington units) is low (less than a
few percent). 
We have then presented a scaling that provides a simple
method to estimate the kinetic energy output of black holes growing at
sub-Eddington rates. Using such a scaling, we have derived the kinetic
luminosity function of AGN and its redshift evolution. Overall, we
show conclusively that the kinetic power output of the low-$\dot m$ mode of
SMBH growth has a very different redshift dependency from the radiative
one, and matches the phenomenological requirements put forward by
semi-analytic schemes of galaxy formation: it is in fact more
effective at late times (and for the more massive systems).
Our results suggest that the so-called ``radio mode'' of AGN feedback
is simply a jet-dominated {\it accretion mode}, and that its physical
and evolutionary properties are dictated by the physics of accretion
in the vicinity of the SMBH.

%\begin{acknowledgments}
%\end{acknowledgments}


\begin{thebibliography}{}

\bibitem[Allen, et al. (2006)]{all06}
  {Allen, S.W., Dunn, R.J.H., Fabian, A.C., Taylor, G.B., Reynolds,
  C.S.} 2006,\textit{MNRAS}, 372, 21 

\bibitem[Bower, et al. (2006)]{bow06}
{Bower, R., Benson, A.J., Malbon, R., Helly, J.C., Frenk, C.S., Baugh,
  C.M., Cole, S., Lacey, C.G.} 2006, \textit{MNRAS}, 370, 645

\bibitem[Croton, et al. (2006)]{cro06}
{Croton, D.J., Springel, V., White, S.D.M., De Lucia, G., Frenk, C.S.,
  Gao, L., Jenkins, A., Kauffmann, G., Navarro, J.F., Yoshida, N.}
2006, \textit{MNRAS}, 365, 11

\bibitem[De Zotti, et al. (2005)]{dz05}
{De Zotti, G., Ricci, R., Mesa, D., Silva, L., Mazzotta, P., 
Toffolatti, L., Gonz\'alez-Nuevo, J.} 2005, \textit{A\&A}, 431, 893

\bibitem[Di Matteo, et al. (2005)]{dm05}
{Di Matteo, T., Springel, L., Hernquist, L.} 2005, \textit{Nature},
433, 604


\bibitem[Falcke, K\"ording \& Markoff (2004)]{fkm04}
{Falcke, H., K\"ording, E. \& Markoff} 2004, \textit{A\&A}, 414, 895

\bibitem[Fender, Gallo \& Jonker (2003)]{fgj03}
{Fender, R.P., Gallo, E. \& Jonker, P.} 2003, \textit{MNRAS}
(Letters), 343, L99

\bibitem[Filho, Barthel \& Nagar (2006)]{fbn06}
{Filho, M.E., Barthel, P.D., Ho, L.C.} 2006, \textit{A\&A}, 451, 71

\bibitem[Gallo, et al. (2006)]{gal06}
{Gallo, E., Fender, R.P., Miller-Jones, J.C.A., Merloni, A., Jonker,
  P.G., Heinz, S., Maccarone, T., van der Klis, M.} 2006,
\textit{MNRAS}, 370, 1351

\bibitem[Heinz \& Grimm (2005)]{hg05}
{Heinz, S. \& Grimm, H.-J.} 2005, \textit{ApJ}, 633, 384 

\bibitem[Marconi \& Hunt (2003)]{mh03}
     {Marconi, A. \& Hunt, L.K.} 2003,
     \textit{ApJL} 589, L21

\bibitem[Marconi, et al. (2004)]{mar04}
{Marconi,A., Risaliti,G., Gilli,R., Hunt,L., Maiolino,R.,
  Salvati,M.} 2004, \textit{MNRAS}, 351, 169

\bibitem[Merloni, Heinz \& Di Matteo (2003)]{mhd03}
{Merloni, A., Heinz, S. \& Di Matteo, T.} 2003, \textit{MNRAS}, 345,
1057 [MHD03]

\bibitem[Nulsen, et al. (2006)]{nul06}
{Nulsen, P.E.J., et al.} 2006, Proc. of \textit{Heating and Cooling in
  clusters of galaxies}, H. Boehringer, P. Schuecker, G.W. Pratt, 
A. Finoguenov (eds.). astro-ph/0611136


\bibitem[Rafferty, et al. (2006)]{raf06}
{Rafferty, D.A., Mc Namara, B.R., Nulsen, P.E.J., Wise, M.W.} 2006,
\textit{ApJ}, in press. astro-ph/0605323

\bibitem[Rees, et al. (1982)]{ree82}
{Rees, M.J., Begelman, M.C., Blandford, R.D., Phinney, E.S.} 1982,
\textit{Nature}, 295, 17

\bibitem[Urry \& Padovani (1991)]{up91}
{Urry, M. \& Padovani, P.} 1991, \textit{ApJ}, 371, 60

\bibitem[Yu \& Tremaine (2002)]{yt02}
{Yu, Q. \& Tremaine, S.} 2002, \textit{MNRAS} 335, 965

\end{thebibliography}
\end{document}